\documentclass[a4paper,11pt]{article}
\usepackage{pos}
\usepackage{braket}

\title{First steps towards the quantum simulation of jet quenching
}

\author*[a]{João Barata}
\author[a]{Carlos Salgado}

\affiliation[a]{  Instituto Galego de F\'{i}sica de Altas Enerx\'{i}as (IGFAE), Universidade de Santiago de Compostela, \\
E-15782 Galicia, Spain}

\emailAdd{joaolourenco.henriques@usc.es}
\emailAdd{carlos.salgado@usc.es}

\abstract{The leading order $\alpha_s$ effect in jet quenching corresponds to the broadening of the jet's transverse momentum, due to the multiple interactions with the underlying medium. A complete understanding of momentum broadening is critical for the success of jet quenching phenomenology.

In this talk, we introduce a strategy to quantum simulate single particle momentum broadening in a QCD background medium. We argue that it is, in principle, possible to extract the jet quenching parameter $\hat{q}$ from such an algorithm. More importantly, this corresponds to the first step towards simulating full medium induced parton showers, which is far beyond the capabilities of classical computers.}

\FullConference{%
  *** The European Physical Society Conference on High Energy Physics (EPS-HEP2021), ***\\
  *** 26-30 July 2021 ***\\
  *** Online conference, jointly organized by Universität Hamburg and the research center DESY ***
}

\def\cA{{\cal A}}

\def\cT{{\cal T}}

\def\cH{{\cal H}}

\newcommand{\secn}[1]{Section~1}
\newcommand{\appn}[1]{Appendix~1}

\long\def\comment#1{ }

\def\and{\quad\text{and}\quad}

\def\0{{\boldsymbol 0}}
\def\p{{\boldsymbol p}}
\def\l{{\boldsymbol l}}

\def\n{{\boldsymbol n}}
\def\x{{\boldsymbol x}}
\def\y{{\boldsymbol y}}
\def\X{{\boldsymbol X}}

\def\0{{\boldsymbol 0}}

\def\P{{\boldsymbol P}}
\def\X{{\boldsymbol X}}

\begin{document}
\maketitle

\section{Introduction}
One of the strongest evidences for the production of a quark gluon plasma in heavy ions collisions is the suppression of high transverse momentum jets' yield. This phenomenon was first anticipated by Bjorken~\cite{Bjorken:1982tu} and is nowadays broadly referred to as jet quenching. Such modifications to the jets properties, when compared to a vacuum benchmark, arise from the interactions of the jets' constituents with the underlying QCD medium. At eikonal accuracy, where energy suppressed terms are neglected, the jet's partons evolve along a common light cone direction at fixed transverse positions with respect to the jet core. As a result, only their individual color fields can be modified by the medium. At this order, all partons satisfy the on-shell relation and thus there is no radiative energy loss, and the only possible phenomenological effect that can be studied corresponds to the broadening of the jet's momentum.  Although single particle momentum broadening can be easily computed  using classical methods, simulating it in quantum computers is a natural starting point for the future simulation of fully embedded jets. For these, multi-particle quantum interferences are crucial. However, tacking such contributions into account is particularly hard to perform in classical computers and thus one hopes that the so-called \textit{quantum advantage} might become important to achieve such a goal in the future. 

We introduce a simple quantum algorithm to simulate the evolution of a single parton in the presence of a QCD background. The strategy followed is a quantum analog of previous classical approaches~\cite{Li:2020uhl} and can be extended to include in-medium radiation.

\section{Theoretical set up}
In this section, we first detail the set up to describe the evolution of a single hard parton in a stochastic QCD background field. From this we will extract the effective Hamiltonian that will then be used to construct the time evolution operator to be implemented in terms of a quantum circuit. Secondly, we give a broad overview of the quantum simulation algorithm, which is used to mimic the evolution of the single particle in the quantum computer. For more details see~\cite{Barata:2021yri}.

\subsection{Parton evolution in a QCD background}
\begin{figure}[h!]
  \centering
  \includegraphics[width=0.5\textwidth]{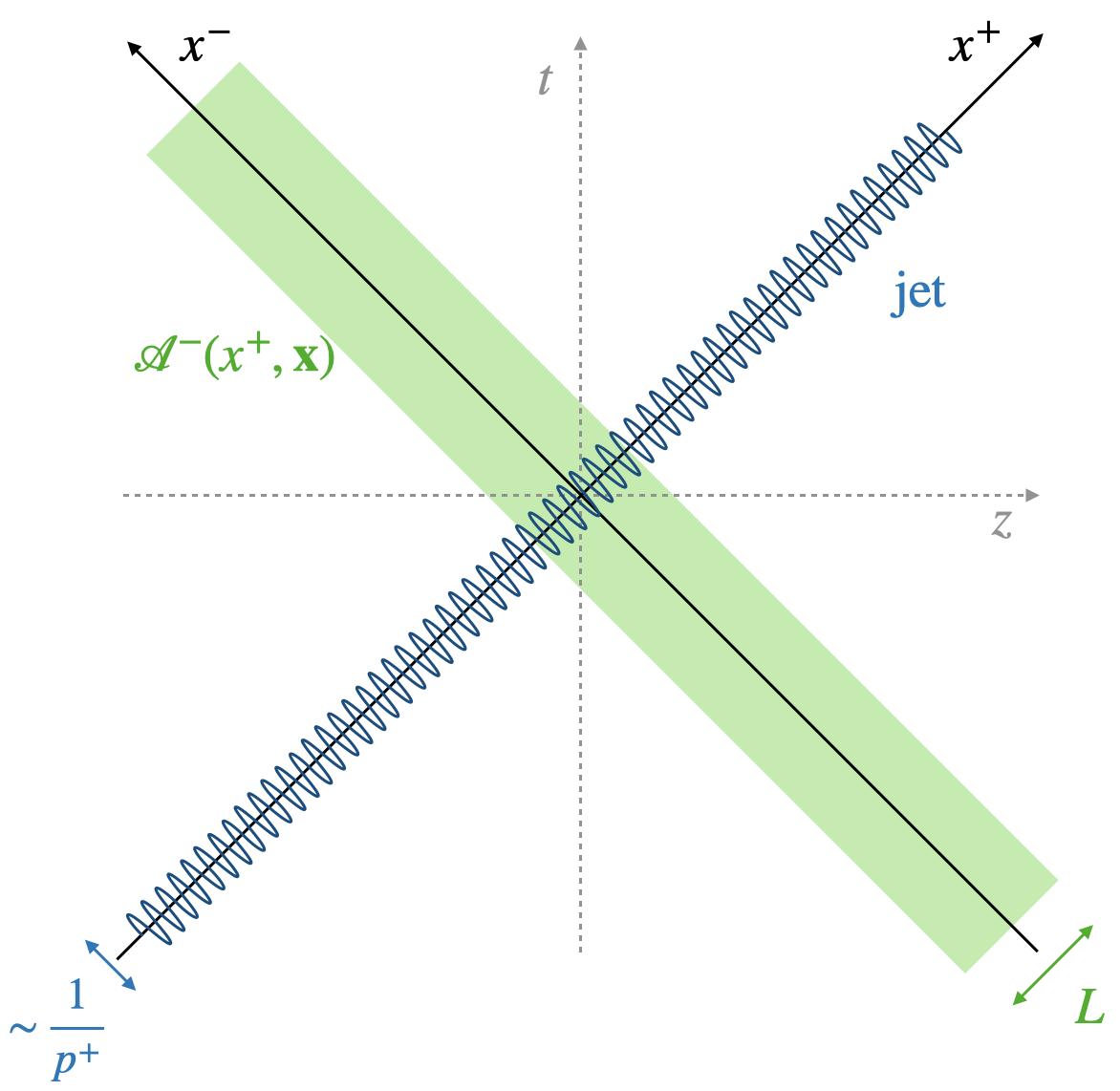}
  \caption{Spacetime picture for a highly boosted parton (jet) evolving in the presence of a QCD background. The medium is assumed to have a finite length $L$ along $x^+$ and $p^+=\omega$.}\label{fig:spacetime}
\end{figure}

We consider a single parton with momentum $p\equiv(\omega,\p,p^-)=((p^0+p^z)/2,\p,p^0-p^z)$ in light cone coordinates. Here we must assume that $\omega\gg |\p|\gg T $, where $T$ denotes the medium temperature and $\omega\approx p^0$ is assumed to be the large component of the momentum of the right-moving particle. The quark gluon plasma background corresponds to a highly occupied state and thus it admits a classical (stochastic) description in terms of a field $\cA^\mu(x)=(\cA^+(x),\cA_\perp(x),\cA^-(x))$. In the light-cone gauge, where $\cA^+=0$, the only non-vanishing field component is $\cA^-$~\cite{Blaizot:2008yb}.  In addition, because the initial parton is highly boosted, the spread of its wavefunction in $p^-$ must be very small. As a consequence, the spacetime dependence of the field can be simplified to be $\cA^-(x^+,\x)\equiv \cA^-(x^+,\x,0)$. Also, in the highly boosted and collinear set up here considered, one can use that $x^0=t\sim x^3$, so that $x^+\sim t$ can be viewed as the time variable of the problem, while all the dynamics of the single parton system are restricted to the two-dimensional transverse plane. Fig.~\ref{fig:spacetime} summarizes the previous discussion on the spacetime picture of the problem.

In the case where one neglects the production of in-medium radiation (i.e. suppresses all $\mathcal{O}(\alpha_s)$ terms), the full theory is captured by the in-medium propagator $G(t,\x;0,\y)$, which obeys~\cite{Blaizot:2012fh}
\begin{equation}\label{eq:Sch_G}
  \left(i\partial_t+\frac{\partial^2_\x}{2\omega}+g\cA^-(t,\x)\cdot T\right)G(t,\x;0,\y)=i\delta(t)\delta(\x-\y) \, .
  \end{equation}
This describes the propagation of a scalar particle between the transverse position $\y$ at time $x^+=0$ and the position $\x$ at time $x^+=t$. In particular, it implies that the corresponding Hamiltonian is that of a non-relativistic particle~\cite{Blaizot:2015lma,Barata:2021yri,Brodsky:1997de}
\begin{equation}\label{eq:Hamiltonian}
\cH(t)=\frac{\p^2}{2\omega}+g\cA^-(t,\x)\cdot T=\cH_K + \cH_\cA(t)   \, .
\end{equation}
This will be the operator to be implemented in terms of quantum circuits. We note that that the eikonal limit mentioned above corresponds to the case where $ \frac{\p}{\omega}\to 0$, and thus only $\cH_{\cA}$ survives. Indeed, this term is diagonal in a position basis and thus only gives rise to a phase factor. Also, in what follows we will assume that the particle is in the singlet color representation, such that $T=1$. The generalization of the algorithm to other color representations is discussed in~\cite{Barata:2021yri}.

\subsection{The quantum simulation algorithm}
The quantum simulation of complex quantum systems by using simpler and controllable ones was first envisioned by Feynman~\cite{Feynman:1981tf} in the 1980s. In its essence, the quantum simulation algorithm consists in solving the Schrodinger equation for a given Hamiltonian by manipulating the natural dynamics of a given quantum system such that it mimics the evolution of a system of interest. In the case where the controllable system is a digital quantum computer, i.e. a collection of interconnected (idealized) $1/2$-spins which admit the application of an universal set of fundamental quantum gates, the algorithm can be broadly summarized as follows~\cite{NielsenChuang}:
\begin{enumerate}
  \item \textbf{Input}: One needs to provide a Hamiltonian $\cH$ and specify the associated Hilbert space. The initial wavefunction of the system $|\psi_0\rangle$ also has to be provided. In addition, for the present problem, the background field is assumed to be a stochastic variable. Thus, one also has to provide an ensemble of $m$ field configurations with the associated probability distribution.
  \item \textbf{Digitization}: Next, it is necessary to map the degrees of freedom of the physical system to the qubits available in the computer. This should also include the qubit representation of the initial wavefunction. 
  \item \textbf{Initial state preparation}: Given the qubit representation of $|\psi_0\rangle$, one needs to prepare a circuit that generates this state in the quantum computer efficiently and accurately. In this step, one assumes that a fiducial state can always be prepared by the computer.
  \item \textbf{Time evolution}: Given the discretization/digitization strategy used, one has to represent the time evolution operator, $U=\exp(-i\int dt \, \cH(t))$, in terms of basic quantum gates. Then one applies these set of gates to the initial state, so that the final state of the system is prepared.
  \item \textbf{Measurement}: Due to the special character of measurement in Quantum Mechanics, one can not efficiently measure the full final wavefunction in general. Instead, measurement protocols have to be implemented allowing one to access  particular information about the system requiring only a reasonable number of measurements.
\end{enumerate}

\section{Quantum simulating single parton broadening}
\begin{figure}[h!]
  \centering
  \includegraphics[width=0.8\textwidth]{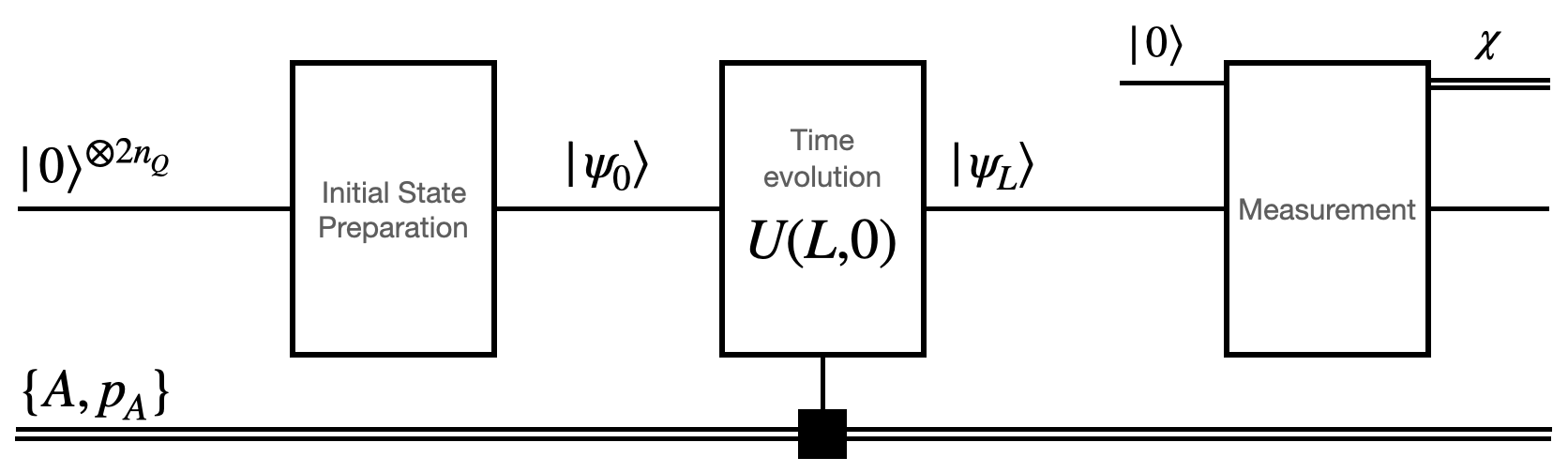}
  \caption{Overview of the quantum circuit. Single lines denote quantum channels while double lines denote classical ones. Above each line we detail the state being store in the circuit. The $\blacksquare$ denotes that the time evolution gates parameters are to be determined from the gauge field.}
  \label{fig:1}
\end{figure}

The quantum simulation algorithm can be immediately applied to simulate the evolution of a single parton inside a medium; the respective quantum circuit is detailed in Fig.~\ref{fig:1}. Note that the bottom part of the diagram denotes a classical memory where the information about the gauge field is stored to then be used by the quantum circuit above.

Following the steps detailed in the previous section, the input to the system will consist on the Hamiltonian given in Eq.~\ref{eq:Hamiltonian}, $m$ field configurations for the QCD background and we shall consider the initial wavefunction to be that of a particle with light-cone energy $\omega$ and initial traverse momentum $\p=\0$. The Hilbert space under consideration corresponds that of a single non-relativistic particle in two dimensions; thus it is natural to describe the system either in a position basis, $|\x\rangle$, or in a momentum on, $|\p\rangle$, which are naturally related by a Fourier transform.

To represent the system in terms of qubit degrees of freedom, we discretize the system in a two dimensional lattice. Any position state can be written as $|\x\rangle=|a_s\n \rangle$, where $a_s$ the spatial lattice spacing and $\n=(n_1,n_2)$ is a dimensionless vector with each component taking values between $0$ and $N_s-1$, with $N_s$ the number of lattices sites per dimension. An equivalent lattice exists in momentum state, related to this one by a discrete Fourier Transform; see~\cite{Barata:2021yri,Klco:2018zqz, Barata:2020jtq} for more details. 

Rewriting the Hamiltonian, $H=\cH a_s$, in terms of position and momentum operators acting on the dimensionless basis introduced
\begin{equation}\label{eq:dimless_H}
  H=\frac{\P^2}{2 E} + g A(t,\X)\cdot T = H_K + H_A(t)   \, , 
\end{equation}
we obtain that evolution operator simply reads 
\begin{equation}\label{eq:U}
  U(L^\prime,0)  \equiv   \cT \exp\left[-i\int_0^{L^\prime} dt\, H(t)\right] \, ,  
  \end{equation}
where $L^\prime\equiv L/a_s$ is the dimensionless medium length. To implement Eq.~\ref{eq:U} in terms of basic quantum gates we use a simple product formula (see~\cite{Poulin_2011, Berry_2015,Berry_2020,Wiebe_2010,Berry_2014} for more discussion on this procedure for time dependent Hamiltonians), which leads to
  \begin{equation}\label{eq:U_TS}
  U(L^\prime,0)\approx   \prod_{k_t=1}^{N_t}    \left\{ \exp\left[-i H_K \frac{L^\prime}{N_t}\right]\exp\left[-i H_A\left(k_t\cdot \frac{L^\prime}{N_t}\right) \frac{L^\prime}{N_t}\right]\right\} \equiv \prod_{k_t=1}^{N_t}    \left\{ U_K(\varepsilon_t) U_A(k_t\cdot \varepsilon_t,\varepsilon_t)\right\} \, .
  \end{equation}
This decomposition of the evolution operator essentially consists in slicing the evolution time $L'$ into $N_t$ steps, where in each step one can use that $e^{X+Y}\approx e^X e^Y$ for $X$ and $Y$ non-commuting operators. Thus, the error associated with this scheme is of $\mathcal{O}(\varepsilon_t^2)$ with $\varepsilon_t=L'/N_t$. In the particular case of the Hamiltonian given in Eq.~\ref{eq:dimless_H}, it is convenient to split it, at each step in the time evolution, in terms of a term governed by $H_K$ and a term containing the gauge field. The reason for this separation is because the kinetic term is diagonal in momentum space, i.e. when acting on a state $|\p\rangle$ it only gives a phase
\begin{equation}
  U_K(\varepsilon_t) |\p\rangle= \exp\left(-i\frac{\varepsilon_t}{2E}\p^2\right) |\p\rangle\, ,       
  \end{equation}
while $H_A$ is diagonal in a position basis
\begin{equation}\label{eq:UA}
  U_A(k_t\cdot \varepsilon_t,\varepsilon_t) |\x\rangle= \exp(-ig\varepsilon_t A(k_t\cdot \varepsilon_t,\x)) |\x \rangle\, .        
  \end{equation}
Thus, if in between the application of the two evolution operators one performs a (quantum) Fourier Transform~\cite{NielsenChuang}, the circuit can be efficiently implemented.

\begin{figure}[h!]
  \centering
  \includegraphics[width=0.8\textwidth]{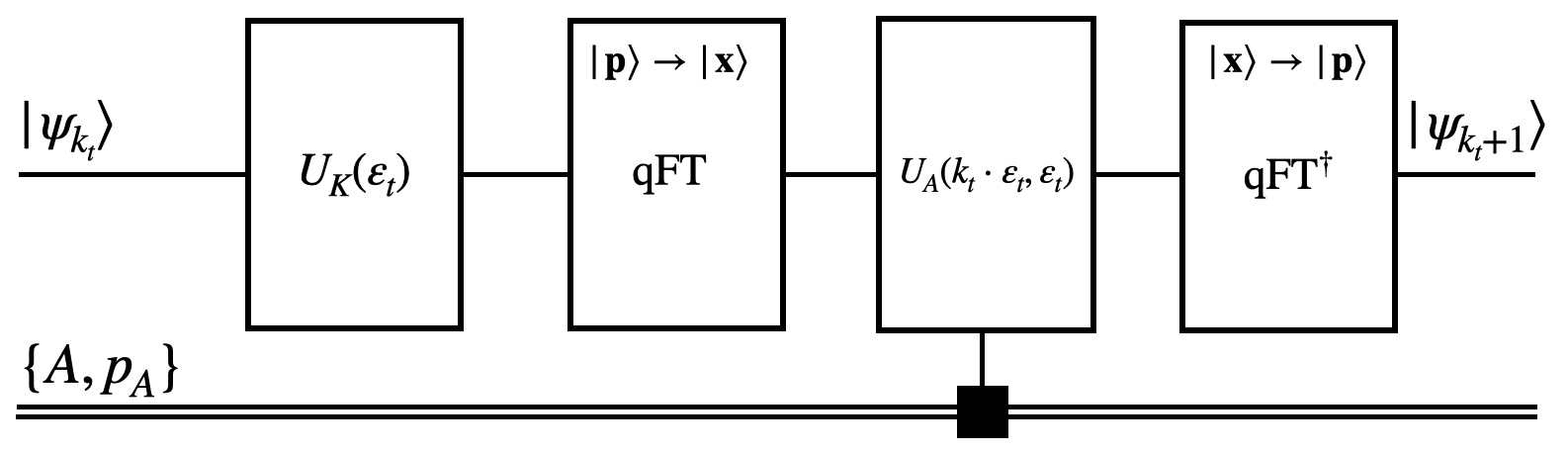}
  \caption{Implementation of the $k_t^{\rm th}$ time step, as indicated in Eq.~\ref{eq:U_TS}.}
  \label{fig:2}
\end{figure}

The decomposition of $U_K$ and $U_A$ can be done in a straightforward manner. For $U_K$ one just needs to implement an algorithm which gives a phase to the ket state as a function of the state value. A simple strategy to realize such an operation was introduced in~\cite{Zalka_1998} and more details on its application can be found in~\cite{Barata:2021yri}. For the gauge term, and assuming that one has transformed from the momentum to the position basis, the implementation requires that a classical memory with the values of the field at time $k_t \varepsilon_t$ is provided for all lattice points. Although this in general a classically costly implementation, we notice that in general one provides a statistical model for the field, which makes its computation quite efficient. Given these field values, one only needs a diagonal operator where each entry contains the field value at a different position; such an operator can be constructed either by classically solving a system of $N_s^2$ linear equations or by multiple applications of controlled quantum gates; see~\cite{Barata:2021yri,NielsenChuang} for some details. A single step of this algorithm is summarized in Fig.~\ref{fig:2}.

After iterating $N_t$ times the above single-step circuits on the initial wave function, one generates the final state $|\psi_L\rangle$. As mentioned in the previous section, given this state one needs to implement a protocol to efficiently extract the relevant information for the wavefunction. In the present work, we are interest in extracting the jet quenching parameter $\hat q$, which is trivially related to the average momentum squared accumulated in the medium. 

\begin{figure}[h!]
  \centering
  \includegraphics[width=0.8\textwidth]{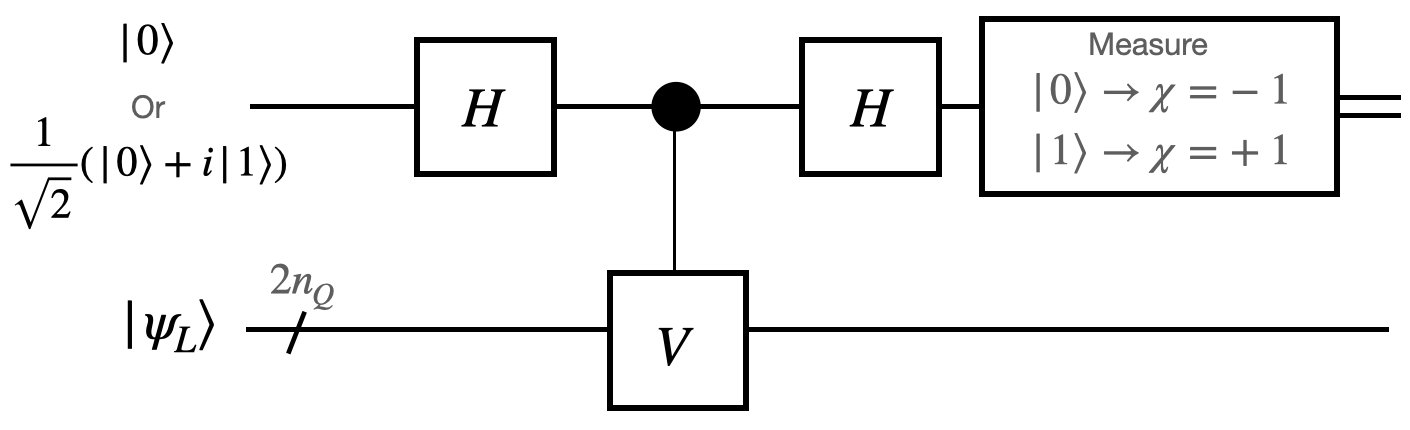}
  \caption{Circuit representation of the measurement strategy.}
  \label{fig:meas}
\end{figure}

In order to this, we apply the Hadamard test as detailed in Fig.~\ref{fig:meas}. We introduce an ancillary qubit which can be either in the state $|0\rangle$ or in the state $|0\rangle+i|1\rangle$. Then we apply the Hadamard transformation~\cite{NielsenChuang} followed by a controlled application of an operator $V$. By controlled we mean that $V$ is only applied to the final state $|\psi_L\rangle$ if the ancilla qubit is in the state $|1\rangle$. Finally, one reverts the Hadamard transform and measures the ancilla. We associate the outcome of such a measurement to a classical random variable $\chi$ which takes value $\chi=-1$ if one measures the state $|0\rangle$ and $\chi=-1$ is one observes the state $|1\rangle$.

It is then easy to show that for any $V$ 
\begin{equation}
\langle \chi \rangle_{\rm QM}\equiv\bra{\psi_L} V+V^\dagger \ket{\psi_L}= \Re    \bra{\psi_L} V \ket{\psi_L} \, . 
\end{equation}
if the ancilla is in the $\ket{0}$ state. If the ancilla is prepared in the state $1/\sqrt{2}(\ket{0}+i\ket{1})$, then
\begin{equation}
\langle \chi \rangle_{\rm QM}= \Im    \bra{\psi_L} V \ket{\psi_L} \, ,
\end{equation}
which gives access to the real and imaginary parts of unitary operator $V$. In the case when $V$ is diagonal in the momentum basis, i.e. $V=V_\alpha=\exp(i\alpha \P^2)$, with $\alpha$ a free numerical parameter, we get that 
\begin{equation}
\Re    \bra{\psi_L} V_\alpha \ket{\psi_L} =\langle \cos (\alpha \P^2) \rangle_{\rm QM} \, ,
\end{equation}
and 
\begin{equation}
\Im   \bra{\psi_L} V_\alpha \ket{\psi_L} = \langle \sin (\alpha \P^2) \rangle_{\rm QM}   \,,
\end{equation}
One can imagine that $\alpha$ can be taken sufficiently small such that to linear order
\begin{equation}
\langle e^{i\alpha \P^2}\rangle_{\rm QM}\approx 1 +i \frac{\alpha}{a_d^2} \hat{q}L\to \langle \sin(\alpha \P^2)\rangle_{\rm QM} \approx \frac{\alpha}{a_d^2} \hat{q}L \, .
\end{equation}
where in the right hand side we have related the expectation value of the operator $V$ to the jet quenching parameter, while the left hand side is directly obtained from the quantum computer. Higher order moments of the underlying distribution can be obtained using the same method but allowing $\alpha$ to be larger~\cite{Barata:2021yri}.

\section{Conclusions and outlook}
In this work we have introduced a simple hybrid quantum strategy to extract the jet quenching parameter. Although the current proposal can not outmatch classical approaches, it offers a road for the future quantum simulation of embedded jets.

In the future, it would be important to have a first-time quantum computation of single particle evolution inside the medium. Due to the necessity to discretize space and the limitations of current hardware, such simulations might still not be able to compete with their classical counterparts. However, we note that the study of color flow in the medium, which requires far less quantum resources, should already be able to reveal important information about jet evolution. Another important goal to pursue would be to extend the present approach to include the production of gluonic radiation.

\bibliographystyle{elsarticle-num}
\bibliography{Lib.bib}  

\end{document}